\documentclass[twocolumn]{aastex631}

\usepackage{amsmath}
\usepackage{multirow}
\usepackage{xspace}
\usepackage{tabularx}

\newcommand{\teff}{\ensuremath{T_{\rm eff}}\xspace}

\newcommand{\kmsec}{km~s$^{\rm -1}$}
\newcommand{\masyr}{mas~yr$^{\rm -1}$}

\newcolumntype{I}{@{\extracolsep{\fill}}c}

\definecolor{forestgreen}{HTML}{228B22}
\definecolor{urlblue}{HTML}{000000}

\newcommand{\Gaia}{{\it Gaia}\xspace}
\newcommand{\Latte}{\textit{Latte}\xspace}
\newcommand{\Fire}{\textsc{Fire}\xspace}
\newcommand{\Ananke}{\code{Ananke}}

\mathchardef\mhyphen="2D

\newcommand{\roughly}{\ensuremath{ {\sim}\,} }

\newlength{\dhatheight}

\newcommand{\code}[1]{\texttt{#1}\xspace}

\newcommand{\unit}[1]{\ensuremath{\mathrm{\,#1}}\xspace}

\newcommand{\km}{\unit{km}}

\newcommand{\Msun}{\unit{M_\odot}}

\newcommand{\e}{\unit{e^{-}}}

\newcommand{\secref}[1]{Section~\ref{sec:#1}}

\newcommand{\bandvar}[2][]{%
  \ifthenelse{\isempty{#1}}{\var{#2}}{\var{#2\_#1}}%
}

\newcommand{\var}[1]{\ensuremath{\texttt{\MakeUppercase{#1}}}\xspace}

\providecommand\physrep{\ref@jnl{Phys.~Rep.}}%
\providecommand\apjs{\ref@jnl{ApJS}}%
\providecommand{\jcap}{\ref@jnl{JCAP}}%

\shorttitle{Ananke DR3}
\shortauthors{Nguyen, Ou, et al}

\graphicspath{{./}{figures/}}

\begin{document}

\title{Synthetic \Gaia DR3 surveys from the FIRE cosmological simulations of Milky-Way-mass galaxies}

\author[0000-0001-6189-8457]{Tri~Nguyen}
\affiliation{Department of Physics and MIT Kavli Institute for Astrophysics and Space Research, \\
77 Massachusetts Avenue, Cambridge, MA 02139, USA}
\affiliation{The NSF AI Institute for Artificial Intelligence and Fundamental Interactions,\\
77 Massachusetts Avenue, Cambridge MA 02139, USA}

\author[0000-0002-4669-9967]{Xiaowei~Ou}
\affiliation{Department of Physics and MIT Kavli Institute for Astrophysics and Space Research, \\
77 Massachusetts Avenue, Cambridge, MA 02139, USA}\footnote{Authors Nguyen and Ou have contributed equally to this work, and therefore this paper should be cited as Nguyen, Ou, et al. (2023)}

\author[0000-0001-5214-8822]{Nondh~Panithanpaisal}
\affiliation{Department of Physics \& Astronomy, University of Pennsylvania, 209 S 33rd St., Philadelphia, PA 19104, USA}

\author[0000-0003-2497-091X]{Nora~Shipp}
\affiliation{Department of Physics and MIT Kavli Institute for Astrophysics and Space Research, \\
77 Massachusetts Avenue, Cambridge, MA 02139, USA}

\author[0000-0003-2806-1414]{Lina~Necib}
\affiliation{Department of Physics and MIT Kavli Institute for Astrophysics and Space Research, \\ 
77 Massachusetts Avenue, Cambridge, MA 02139, USA}
\affiliation{The NSF AI Institute for Artificial Intelligence and Fundamental Interactions,\\
77 Massachusetts Avenue, Cambridge MA 02139, USA}

\author[0000-0003-3939-3297]{Robyn~Sanderson}
\affiliation{Department of Physics \& Astronomy, University of Pennsylvania, 209 S 33rd St., Philadelphia, PA 19104, USA}
\affiliation{Center for Computational Astrophysics, Flatiron Institute, 162 5th Ave., New York, NY 10010, USA}

\author[0000-0003-0603-8942]{Andrew~Wetzel}
\affiliation{Department of Physics \& Astronomy, University of California, Davis, CA 95616, USA}

\correspondingauthor{Tri Nguyen and Xiaowei Ou}
\email{tnguy@mit.edu, xwou@mit.edu}

\begin{abstract}

The third data release (DR3) of \Gaia has provided a five-fold increase in the number of radial velocity measurements of stars, as well as a stark improvement in parallax and proper motion measurements. 
To help with studies that seek to test models and interpret \Gaia DR3, we present nine \Gaia synthetic surveys, based on three solar positions in three Milky Way mass galaxies of the \Latte suite of the \Fire-2 cosmological simulations. 
These synthetic surveys match the selection function, radial velocity measurements, and photometry of \Gaia DR3, adapting the code base \Ananke, previously used to match the \Gaia DR2 release in  Sanderson et al. 2020. 
The synthetic surveys are publicly available and can be found at \url{http://ananke.hub.yt/}. 
Similarly to the previous release of \Ananke, these surveys are based on cosmological simulations and thus able to model non-equilibrium dynamical effects, making them a useful tool in testing and interpreting \Gaia DR3. 

\end{abstract}

\keywords{}

\section{Introduction} \label{sec:intro}

The \Gaia mission \citep{gaia16} has revolutionized the study of our Galaxy, the Milky Way (MW). The second data release \citep[DR2;][]{gaia18} provided positions, proper motions, and parallaxes for over one billion stars, including the first kinematic measurements of many stars across the Galaxy. In addition, DR2 included radial velocities for $\roughly 7$ million stars, making it the largest six-dimensional kinematic catalog at the time. This data has enabled the discovery of new merging events, such as the \Gaia Sausage Enceladus \citep{2018MNRAS.478..611B,2018Natur.563...85H}, and the Kraken \citep{2019MNRAS.486.3180K} (see \cite{2020ARA&A..58..205H} for a review), the construction of a new 3D dust map of the MW \citep{2019ApJ...887...93G}, a detailed study of open clusters to unveil the history of the Galactic disk \citep{2018A&A...618A..93C}, an accurate measurement of the circular velocity of the Galaxy \citep{2019ApJ...871..120E}, and detailed studies of the fine resonances of the MW disk \citep[see e.g.][]{2018Natur.561..360A}. The third data release \citep[DR3;][]{gaia22, gaia21} builds upon DR2, incorporating 12 months of additional observations, and significantly increasing the catalog of stars with 6D phase space measurements, including radial velocities, to $\roughly 33$ million stars, as well as reducing uncertainties on and increasing the size of the sample of stars with full astrometry. These data have further enabled a deeper understanding of the dynamics of the Galaxy, for example by extending measurements of the circular velocity of the MW to larger distances \citep{2023arXiv230312838O}.

Synthetic catalogs and mock observations generated from cosmological simulations provide a valuable comparison to these rich observations of our own Galaxy. They enable tests of analysis tools and of our ability to recover true properties of our Galaxy from observations. \cite{sanderson20}, hereafter S20, produced nine \Gaia DR2 synthetic surveys of the \Latte suite of simulations \citep{Wetzel:2016, Hopkins:2018}, using the code \Ananke. Such synthetic surveys have been used in many studies involving the dynamics of the MW, for example to estimate the detectability of simulated stellar streams \citep{Shipp:2023}, as a training set for a neural network that built the first accreted star catalog in the MW \citep{2020A&A...636A..75O}, leading to a discovery of a prograde local structure Nyx \citep{2020NatAs...4.1078N}, and as a framework to test the ability of unsupervised machine learning techniques to reproduce the stellar phase space density \citep{2023MNRAS.tmp..853B}.
In this work, we present synthetic \Gaia DR3 surveys based on the same suite of \Latte simulations.

The \Latte simulations first introduced in \citep{Wetzel:2016} are baryonic zoom-in simulations of MW analogs from the Feedback in Realistic Environments (\Fire) project \citep{2015MNRAS.450...53H,Hopkins:2018}. With an initial stellar particle mass resolution of $7070 \Msun$, the \Latte simulations resolve stellar populations down to the masses of individual star clusters. They self-consistently model baryonic processes, including star formation and the metal-enrichment of gas, which is essential for accurately calculating the extinction of observed stars. At the same time, they incorporate the effects of galaxy formation in a cosmological context, including a realistic history of mergers and accretion events.

\Ananke is a framework for producing synthetic surveys based on the \Fire simulations, first presented by S20. Such work is based on Galaxia \citep{2011ApJ...730....3S}, which generated synthetic surveys of the MW based on kinematic distributions and N-body simulations. The framework entails sampling a population of individual stars from simulated star particles, assigning them realistic physical properties, and applying a simple error model to produce mock observations. 
\Ananke has been applied to produce mock observations of other surveys from a range of simulated data sets, such as APOGEE \citep{2022RNAAS...6..125B}, Dark Energy Survey \citep[DES;][]{DES:2005, DES:2016}, and the Rubin Observatory Legacy Survey of Space and Time \citep[LSST;][]{LSST:2019} in \cite{Shipp:2023}.

In this paper, we use \Ananke~to produce synthetic \Gaia DR3 surveys of three MW analogs from the \Latte simulation suite, focusing on the updates to the surveys compared to S20. This paper is organized as follows:
In \secref{sims} we review the simulations and mock catalogs used in this work, in \secref{synthetic_survey} we discuss the synthetic \Gaia DR3 observations, and in \secref{results}, we present the resulting synthetic surveys, comparing them with those from \Gaia DR2. We list the columns of the public release and their definitions in Section~\ref{sec:column_list}, and discuss the use cases and limitations of these synthetic surveys in \secref{limitations}.

\section{Simulations and Mock Catalogs}
\label{sec:sims}

In this section, we outline the different steps to build both a mock catalog and a synthetic survey. We first seek to define these two terms. Generating a \emph{mock catalog} consists of spawning stars from star particles in the initial simulations, where the star particles have a mass on the order of $10^{4}M_{\odot}$, depending on the initial simulation resolution. This process is independent of the target survey. Generating a \emph{synthetic survey} involves incorporating the specifics of a particular survey into the catalog of simulated stars, including, for example, photometric passbands, measurement errors, dust extinction, and the observer's location.

In order to build the new synthetic \Ananke DR3 survey, we use the same three zoom-in simulations of MW-mass galaxies from the \Latte suite of \Fire-2 simulations as in S20 (m12i, m12f, m12m)\footnote{For a brief overview of the simulation and the simulated galaxies used, we refer the reader to Section~2 of S20.}. 
The choice of these specific simulations is motivated by \citet{Wetzel:2016,2018ApJ...869...12S}, which have shown that these simulated galaxies reasonably reproduce a realistic galaxy of MW mass. 
Specifically, they have thin and thick disk geometries, with scale heights, scale radii, gas fractions, etc. that are broadly similar to the MW.

\subsection{Locations of the Sun}

To build a synthetic survey, we must assume the location of the observer, which we define as the solar position in the simulation. The procedure we adopt here for the coordinate transformation and the definition of the local standards of rest (LSRs) remain unchanged from S20, which we briefly summarize here.
We assume that the Sun is at $R_{\odot} = 8.2$ kpc \citep{2016ARA&A..54..529B} in the three simulations\footnote{
This is an appropriate approximation given that these simulations have comparable scale heights and radii to the MW. 
}, and define the principle axes based on the moment of inertia tensor of the youngest stars (with ages $<$ 1 Gyr) located within $R_{\odot}$. 

The three positions of the Sun in each galaxy are chosen to be evenly distributed in azimuthal angle, and at vertical distance $Z_{\odot} = 0$ kpc. We define the velocity of LSR as the median velocity of the star particles within 200 pc of the solar position. We summarize the positions and velocities of the LSRs in Table~\ref{tab:lsr}, which matches Table 4 of S20. 

\begin{table*}
\centering
\caption{The coordinates of each LSR as shown in Table 4 of S20.}
\label{tab:lsr}
\begin{tabularx}{0.8\linewidth}{I|III|III|III}
\hline \hline
&
  $x_\mathrm{LSR}$ & $y_\mathrm{LSR}$ & $z_\mathrm{LSR}$ & $v_{x, \mathrm{LSR}}$ & $v_{y, \mathrm{LSR}}$ & $v_{z, \mathrm{LSR}}$ & $v_{R, \mathrm{LSR}}$ & $v_{Z, \mathrm{LSR}}$ & $v_{\phi, \mathrm{LSR}}$ \\
label                                    & (kpc)       & (kpc)    & (kpc)                        & (km/s)       & (km/s) & (km/s)                  & (km/s) & (km/s) & (km/s) \\ \hline
\texttt{m12i-lsr-0} & $0.0$ & $8.2$ & $0.0$ & $224.7092$ & $-20.3801$ &  $3.8954$  & $-17.8$  & $-3.9$ & $224.4$ \\
\texttt{m12i-lsr-1} & $-7.1014$ & $-4.1$ & $0.0$ & $-80.4269$ & $191.7240$ & $1.5039$ & $-24.4$ & $-1.5$ & $210.9$ \\
\texttt{m12i-lsr-2} & $7.1014$ & $-4.1$ & $0.0$ & $-87.2735$ & $-186.8567$ & $-9.4608$  & $22.1$  & $9.5$ & $206.5$ \\
\hline
\texttt{m12f-lsr-0} & $0.0$ & $8.2$ & $0.0$ & $226.1849$ & $14.3773$  & $-4.8906$ & $14.9$ & $4.9$ & $227.9$ \\
\texttt{m12f-lsr-1} & $-7.1014$ & $-4.1$ & $0.0$ & $-114.0351$ & $208.7267$ & $5.0635$ & $-3.4$ & $-5.1$ & $244.3$\\
\texttt{m12f-lsr-2} & $7.1014$ & $-4.1$ & $0.0$ & $-118.1430$ & $-187.7631$ & $-3.8905$ & $-11.4$ & $3.9$ & $227.4$\\
\hline
\texttt{m12m-lsr-0} & $0.0$ & $8.2$ & $0.0$ & $254.9187$ & $16.7901$ & $1.9648$ & $16.2$ & $-2.0$ & $254.7$ \\
\texttt{m12m-lsr-1} & $-7.1014$ & $-4.1$ & $0.0$ & $-128.2480$ & $221.1489$ & $5.8506$ & $2.4$ & $-5.9$ & $252.7$\\
\texttt{m12m-lsr-2} & $7.1014$ & $-4.1$ & $0.0$ & $-106.6203$ & $-232.2056$ & $-6.4185$ & $15.4$ & $6.4$ & $265.3$ \\
\hline \hline
\end{tabularx}
\end{table*}

\subsection{Building a mock catalog}

In this section, we discuss the procedure to build a \emph{mock catalog} by converting the star \textit{particles} from the \textsc{Fire-2} simulations into synthetic stars, leaving the construction of the \emph{synthetic survey} in which we add the correct properties to these synthetic stars as drawn by the survey to Section~\ref{sec:synthetic_survey}. 

Each star particle, with a mass $M_* \sim 7070\,\Msun$, is assumed to represent a population of synthetic stars with a single age and metallicity.
To generate such mock catalogs, S20 used the \textsc{Galaxia} code \citep{2011ApJ...730....3S}. In this work, we adopt the same mock catalogs as in S20, modifying the stellar isochrones used in the generation of stars to the updated \Gaia DR3 isochrones. This modification is performed at Step 2 below, while keeping the masses and the phase-space positions the same as in S20 in Step 1 \& 3.

To build a mock catalog, we perform the following three steps. We will leave a detailed description of the DR3 isochrones to Section~\ref{sec:isochrones}.

\begin{enumerate}
    \item First, we sample the stellar masses of synthetic stars using the initial mass function (IMF) in \cite{Kroupa:2001} until the total mass equals the mass of the particle.\footnote{The number of stars sampled is required to be an integer, while the fraction of the IMF within a subrange of mass is not. Some rounding is assumed, and given that the highest possible stellar mass is still two orders of magnitude lower than the mass of the star particle, in this work, as in S20, we assume that this is a valid approach with a small fractional error.} 
    \item Select the isochrone model that is closest in age and metallicity to the parent particle and obtain stellar properties and \Gaia DR3 passband magnitudes by interpolating across initial stellar mass.
    Only stars with estimated unextincted apparent \Gaia DR2/3 magnitudes of $3 < G < 21$ are kept in the catalog, before applying the full selection function in Section~\ref{sec:selectionfunction}.
    \item Assign the positions in phase space to each star by sampling from an Epanechikov density kernel \citep{epan1969} centered on the parent particle.
    The smoothing kernel is computed from the 6-dimensional phase space coordinates using the \textsc{EnLink} code \citep{2006MNRAS.373.1293S, 2009ApJ...703.1061S}.
    Similarly to S20, we use the nearest 8 neighboring star particles to compute the kernel size. The kernel size taken with respect to two independent smoothing lengths, one for the distances, and one for the velocities. 
    The size is proportional to the geometric mean of the smoothing lengths along each of the three-dimensions. 
    To preserve the dynamic ranges of the different stellar populations and avoiding the oversmoothing of structures from different stellar populations, a kernel is computed for \emph{in situ} stars, which are defined as those formed within $30$~physical kpc of the main galaxy, while a separate kernel is computed for stars formed outside of this radius.
    In addition, we subdivide \emph{in situ} stars into eight age bins corresponding to the populations of the Besan\c{c}on Milky Way model (Table 2.1 of \cite{2012A&A...543A.100R}) and compute a different kernel for each of them.    
    This kernel is not optimized for small-scale structures, and in some cases may introduce unphysical features into substructures such as low mass satellite galaxies and stellar streams. For example, \cite{Shipp:2023} adopted a different kernel (albeit also based on the Epanechikov kernel) but with the 16 neighboring star particles and a kernel size that is inversely proportional to the cube-root of the local density around each parent particle, to properly smooth out stellar streams. 
\end{enumerate}

\section{Synthetic Surveys}
\label{sec:synthetic_survey}

We describe the procedure used to produce the \Ananke DR3 \emph{synthetic surveys}.
As mentioned in Section~\ref{sec:sims}, we use the mock catalogs presented in S20, and apply updated DR3 isochrones (Section~\ref{sec:isochrones}), extinction modeling (Section~\ref{sec:extinction}), observational uncertainty modeling (Section~\ref{sec:errors}), and selection function (Section~\ref{sec:selectionfunction}).

\subsection{Isochrones}
\label{sec:isochrones}

We use updated \Gaia DR3 passbands and isochrones from Padova CMD v3.6\footnote{\url{http://stev.oapd.inaf.it/cgi-bin/cmd}} to generate updated intrinsic \Gaia DR3 magnitudes for stars in the mock catalogs in the $G$, $G_{BP}$, and $G_{RP}$ bands.
The photometric system follows the revised and expanded library described in \citet{chen19}, adopting a revised spectral energy distribution (SED) for Vega from \citet{bohlin20}. 
Two assumptions are made while adopting the isochrones.
First, circumstellar dust is ignored as it mostly affects the bright end of the isochrones, where the grid is the sparsest. Therefore, linear interpolation with the circumstellar dust included creates unphysical features when stars fall between these sparse grid points. 
Second, we remove the isochrone grid points representing white dwarfs, as the transition from the tip of the red giant branch to the white dwarf is not modeled by \textsc{Galaxia}. 
Since \textsc{Galaxia} takes the edge value for magnitudes when a star is outside of the isochrone grid, stars beyond the last non-white dwarf grid point are all assigned the same magnitudes, creating artificial overdensities at the tip of the giant branches in the final sample.
We expect these stars to be potential white dwarfs and flag all affected stars in the final synthetic survey as \texttt{flag\_wd} and recommend removing stars with \texttt{flag\_wd} set to 1 before conducting analysis. 
We expect a minimal effect on the overall completeness of the sample as a result of this treatment. As shown by \citet{fusillo21}, $359,073$ white dwarfs are confidently identified in \Gaia\ DR3, comprising less than $0.05\%$ of the full \Gaia\ sample. Even accounting for the fact that the white dwarf catalog presented by \citet{fusillo21} is less complete in crowded regions near the galactic plane, the total white dwarf count in the actual \Gaia\ catalog is expected to be a tiny fraction of the full sample. Thus, for the synthetic survey, the overall loss in stellar count and impact on sample completeness are expected to be minimal as a result of this cut.

\subsection{Extinction Modeling}
\label{sec:extinction}

We adopt a self-consistent extinction model similar to Section 5.1 of S20, which we briefly describe below.
The \Fire-2 simulations do not resolve the creation and destruction of dust grains, so we assume the line-of-sight extinction by dust traces the metal-enriched gas in the simulations. We calculate the reddening $B-V$ of each star using the metal-weighted column density of hydrogen along the line of sight between the star and the solar position.
The extinction is therefore calculated self-consistently, using the gas and metal distributions of each individual simulated galaxy, and thus accurately captures the spatial structures of the galaxy.
The extinction at 550 nm $A_0$ is calculated using the standard relation, $A_0 = 3.1 E(B-V)$ (\citealp{johnson65}, \citealp{schultz75}, \citealp{whittet80}), and then converted into extinction in the \Gaia DR3 passbands.

Using the coefficient $A_0$ from the \Ananke DR2 mock catalogs, as described in \citet{sanderson20}, we re-calculate the extinction coefficients $A_{G, BP, RP}$ in the \Gaia DR3 passbands. 
We adopt the extinction conversion relation provided by the \Gaia collaboration as part of the auxiliary data for eDR3 to compute $A_{G,RP,BP}$ as functions of $A_{0}$ and the unextincted color $(G_{BP}-G_{RP})$. \footnote{The relationship and coefficients can be downloaded from \url{https://www.cosmos.esa.int/web/gaia/edr3-extinction-law}}.
Specifically, we compute $A_{G, BP, RP} = k_{G, BP, RP} \, A_0$, where $k_{G, BP, RP}$ is a function of $A_{0}$ and $(G_{BP}-G_{RP})$.
Using the extinction coefficients ($A_{G, BP, RP}$), we convert the intrinsic magnitudes interpolated from the isochrones into the extincted intrinsic magnitudes. 
These extincted intrinsic magnitudes are combined with the distance modulus to calculate the true extincted apparent magnitudes. 

Following the recommendation from the \Gaia collaboration, we do not directly apply the extinction law outside of the applicable color range, $-0.06 < G_{\text{BP}} - G_{\text{RP}} < 2.5$. However, excluding stars outside of range introduces an unnatural cut on the $G_{\text{BP}} - G_{\text{RP}}$~distribution. We therefore extrapolate their \Gaia passbands extinction coefficients using the nearest $G_{\text{BP}} - G_{\text{RP}}$ extreme value (i.e. $-0.06$ or $2.5$).

The extinction law is also limited to extinction coefficients ($A_{0}$) in the range from $0.01$ to $20$.
On the low end, since the extinction law returns finite positive values for $k_{G, BP, RP}$, the resulting $A_{G, BP, RP}$ always converge to $0$ as $A_{0}$ goes to 0 and thus the extinction law naturally extrapolates to $A_{0}=0$. 
On the high end, unlike with the color, we do not adopt the extreme value (i.e., $20$) for $A_{0}$ or attempt to approximate the extinction law outside of the applicable $A_{0}$ range. 
Stars with $A_{0}>20$ are not included in the final synthetic survey for two reasons.
Firstly, $A_{0}$ is implicitly related to the distance of the star as the extinction arises from the dust between the star and the observer. If we were to adopt the extreme value for the extinction coefficient for a given star, the reported extincted photometry in the final synthetic survey would be inconsistent with the reported parallax and the underlying dust map. 
Secondly, the extincted apparent magnitudes for the majority of the stars with $A_{0}>20$ are expected to be significantly fainter than the observational limit of the synthetic survey ($G_{\rm{obs}}<21$, as described in Section~\ref{sec:selectionfunction}). We therefore do not expect the cut on $A_{0}>20$ to have a significant impact on the completeness of the final synthetic survey.

\subsection{Error Modeling}
\label{sec:errors}

We construct the photometric error model from the fit \Gaia DR3 photometric uncertainties tool provided by \Gaia DPAC (Data Processing and Analysis Consortium)\footnote{\url{https://www.cosmos.esa.int/web/gaia/fitted-dr3-photometric-uncertainties-tool}}, based on data originally described in \citet{riello21}.
We adopt the astrometric measurement error models from the \texttt{PyGaia} package\footnote{\url{https://github.com/agabrown/PyGaia}}.
The spectroscopic error model is obtained from private communication with the \Gaia collaboration as a function of \teff~and $G_{\rm{RVS}}$. 
We calculate the errors and the error-convolved quantities by randomly sampling from a one-dimensional Gaussian centered on the truth values.
In the final catalog, we report both the truth values and the error-convolved values.

\subsubsection{Photometric Error}
\label{sec:photo_errormodel}

\begin{figure}
\begin{center}
\includegraphics[angle=0,width=0.48\textwidth]{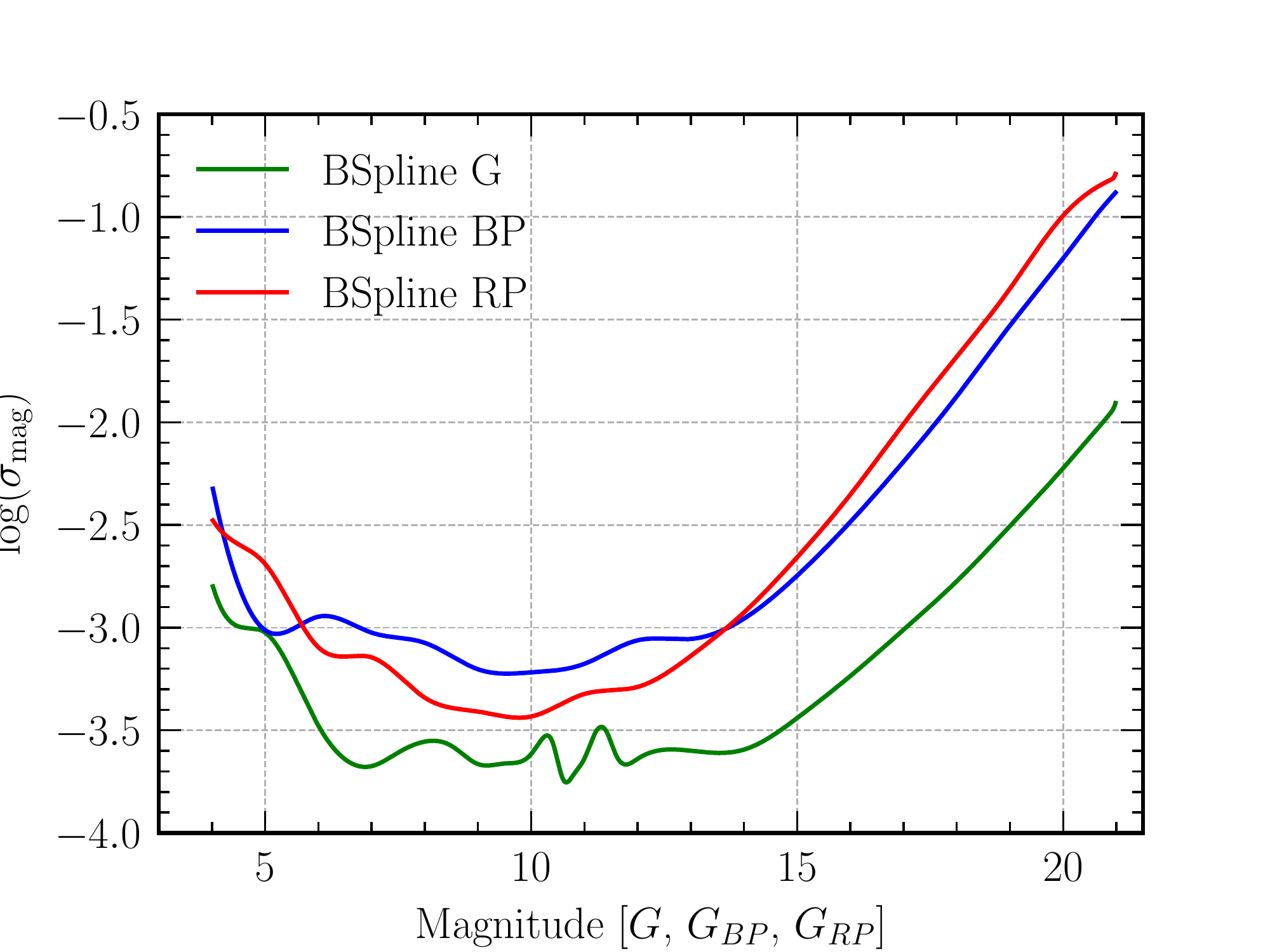}
\end{center}
\caption{
Logarithm of the expected uncertainties for sources with 200(20) observations in $G$($G_{BP}$/$G_{RP}$). 
}
\label{fig:phot_err_model}
\end{figure}

As mentioned, we adopt the photometric uncertainties tool from \Gaia DPAC to calculate the errors in ($G$, $G_{BP}$, $G_{RP}$).
The tool models the median behavior of the real \Gaia (e)DR3 photometric uncertainties in the three \Gaia passbands via cubic B-spline fitting.
The errors in each photometric band are calculated as a function of the band \textit{extincted} magnitudes.
Because the B-spline is restricted to a range $[4, 21]$ in all three bands, we extrapolate the photometric uncertainties of each band using the nearest extreme values (i.e. $4$ or $21$)
In addition, the tool is capable of scaling the fit B-splines with different numbers of observations.
We take, for simplicity, the default number of observations (i.e., 200 for $G$ and 20 for $G_{BP}$/$G_{RP}$) for all stars in our catalogs. 
We show in Figure~\ref{fig:phot_err_model} the errors as a function of extincted magnitude and reproduce Figure~14 of \citet{riello21}.
We note that this error modeling does not take into account systematic effects originating from the properties of the source, e.g., position and color.

\subsubsection{Astrometric Error}
\label{sec:astro_errormodel}

\texttt{PyGaia} models the astrometric errors (i.e. parallax, position, and proper motion) as solely dependent on the apparent $G$ magnitude. 
The position and proper motion errors are returned in the ICRS frame, i.e. in RA and Dec.
To obtain the error-convolved positions and proper motions in Galactic coordinates ($\ell, b$), we calculate the error-convolved ICRS coordinates and apply a coordinate transformation.
Similarly to \Gaia, we do not report the error in the Galactic coordinates. 
The astrometric errors, $\sigma_{\rm X}$, for $X \in (\alpha_\star, \delta, \mu_{\alpha \star}, \mu_\delta)$, can be summarized as follows:
\begin{equation}
    \label{eq:astro_error}
    \sigma_X = c_X \sigma_\varpi, \quad \sigma_\varpi = \sqrt{40 + 800 z + 30 z^2}
\end{equation}
\begin{equation}
    \quad \log_{10}z = 0.4 (\mathrm{max}[G, G_{\mathrm{bright}}] - 15.0),
\end{equation}
where $\sigma_\varpi$ is the parallax error, and $G_\mathrm{bright}=13$.
The coefficients $c_X$ are reported in Table~\ref{tab:astro_error_coeff}. 
Because \texttt{PyGaia} returns the error in RA cos(Dec) $\sigma_{\alpha \star}$, we convert $\sigma_{\alpha \star}$ to the RA error $\sigma_\alpha$ via analytical error propagation.

\begin{deluxetable}{ccccc}
\caption{Coefficients of the astrometric errors in Eq.\ref{eq:astro_error}.}
\label{tab:astro_error_coeff}
\tablehead{
\colhead{$\varpi$}& 
\colhead{$\alpha_\star$}& 
\colhead{$\delta$} & 
\colhead{$\mu_{\alpha\star}$} & 
\colhead{$\mu_\delta$}
}
\startdata
$1.0$ & $0.80$ & $0.70$ & $1.03$ & $0.89$
\enddata
\end{deluxetable}

\subsubsection{Spectroscopic Error}
\label{sec:spec_errormodel}
For spectroscopic measurements, \Gaia DR3 provides radial velocity spectra (with magnitude $G_\mathrm{RVS}$), object classifications, and measured stellar parameters, such as effective temperature, surface gravity, extinction coefficient, and metallicity, in addition to radial velocities. 
Our synthetic survey only provides error-convolved radial velocity measurements. 
For DR3 radial velocities, we first use relationships provided by \citet{sartoretti22} to obtain true \Gaia RVS magnitude, $G_{RVS}$, from $G$ and $G_{RP}$. 
To do so, we use
\begin{equation}
\begin{aligned}
    G_{RVS} - G_{RP} &= a_0 + a_1(G-G_{RP}) \\
    &+ a_2(G-G_{RP})^2 + a_3(G-G_{RP})^3,
\end{aligned}
\end{equation}
where the coefficients are provided in Table~\ref{tab:grvs_coeff}. As for the extinction law extrapolation, we approximate the conversion for stars outside of the applicable range ($-0.15<G-G_{\text{RP}}<1.7$), using the coefficients corresponding to the nearest $G - G_{\text{RP}}$ extreme value (i.e., $-0.15$ or $1.7$). 

\begin{deluxetable}{ccccc}
\caption{Coefficients for color transformation from $G-G_{RP}$ to $G_{RVS} - G_{RP}$.}
\label{tab:grvs_coeff}
\tablehead{
\colhead{$a_0$} & 
\colhead{$a_1$} & 
\colhead{$a_2$} & 
\colhead{$a_3$} & 
\colhead{$G-G_{\text{RP}}$ range} 
}
\startdata
$-$0.0397 & $-$0.2852 & $-$0.0330 & $-$0.0867 & $[-0.15, 1.2]$ \\
$-$4.0618 & 10.0187 & $-$9.0532 & 2.6089 & $[1.2, 1.7]$
\enddata
\end{deluxetable}

\begin{figure}
\begin{center}
\includegraphics[angle=0,width=0.48\textwidth]{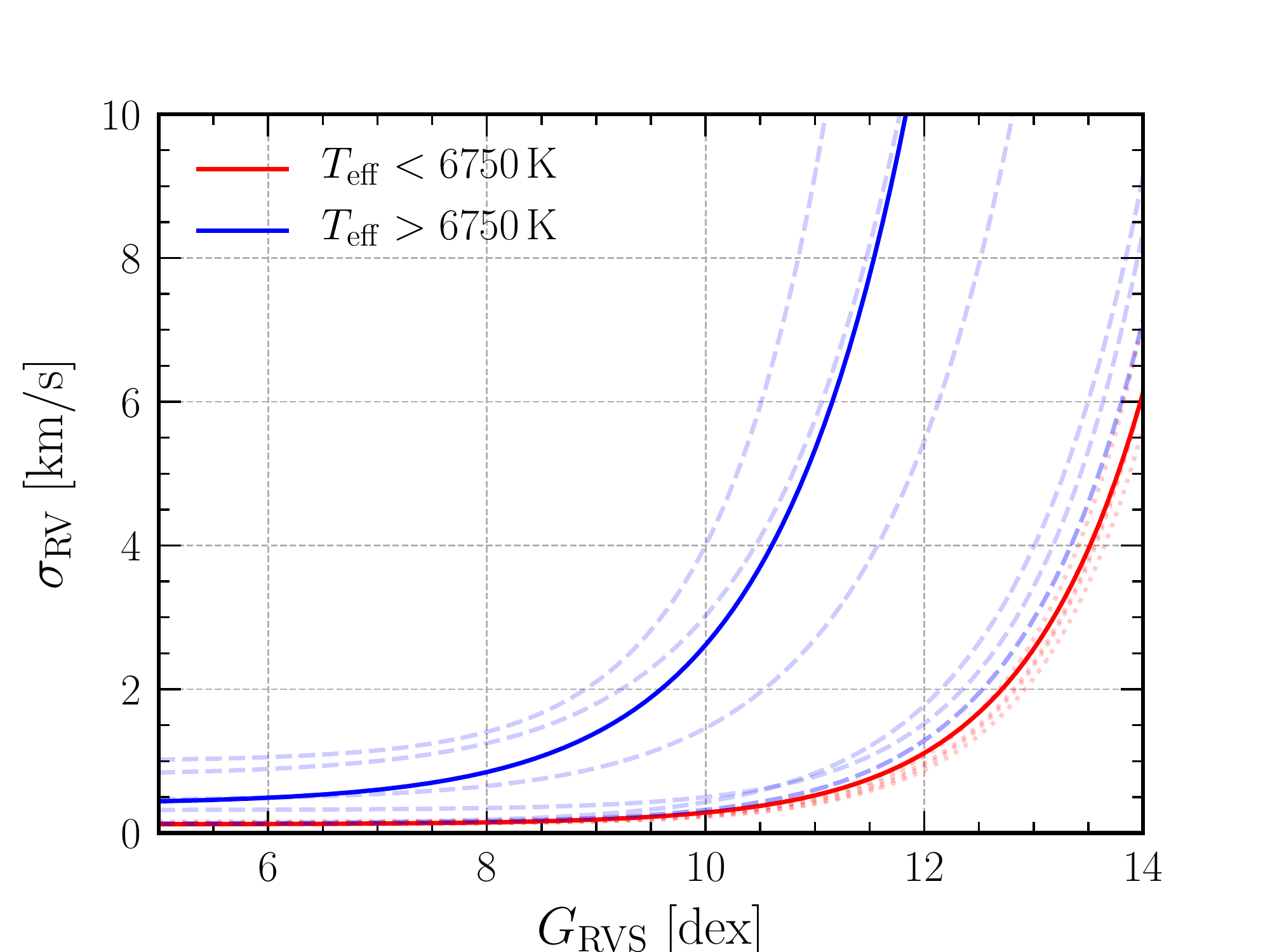}
\end{center}
\caption{
Comparison between the radial velocity error models adopted in this study (solid line) and those provided by the \Gaia~collaboration with the official release of DR3 (dashed/dotted lines). Blue dashed lines represent error estimates for dwarfs, whereas red dotted lines represent error estimates for giants. Our estimate for cooler stars (\teff$< 6750$~K) is largely consistent with the estimates for giants and, similarly, our estimate for warmer stars (\teff$> 6750$~K) with the estimates for dwarfs.
}
\label{fig:spec_err_model}
\end{figure}

The radial velocity uncertainty is fit as a function of $G_{\rm{RVS}}$,
\begin{equation}
\begin{aligned}
    \sigma_{\text{RV}} = \sigma_{\text{floor}} + b \exp{(a(G_{\text{RVS}}-G_{\text{RVS,0}}))}.
\end{aligned}
\end{equation}
The coefficients $a,b$ are fit independently for cooler (\teff$< 6750$~K) and warmer (\teff$> 6750$~K) stars (Table~\ref{tab:rv_err_coeff}), obtained from \citetext{Private Communication, Gaia Collaboration, 2022} prior to the official release of the third data release. Warmer stars generally have a larger error in radial velocities.
While the error modeling for warm stars, as shown in Figure~\ref{fig:spec_err_model}, appears to greatly exceed 10~\kmsec at the very faint end ($G_{\rm{RVS}} \sim 14$), we note that only warm stars with $G_{\rm{RVS}} < 12$ are selected to have a measured radial velocity in the final catalog, as described in more detail in Section~\ref{sec:selectionfunction}.
The maximum radial velocity measurement uncertainties are thus $\roughly 6$~\km/s for cool stars and $\roughly 11$~\km/s for warm stars.  

\begin{deluxetable}{ccccc}
\caption{Coefficients for $\sigma_{\text{RV}}$ as a function of $G_{\text{RVS}}$}
\label{tab:rv_err_coeff}
\tablehead{
\colhead{$\sigma_{\text{floor}}$} & 
\colhead{$a$} & 
\colhead{$b$} & 
\colhead{$G_{\text{RVS,0}}$} & 
\colhead{Applicable range} 
}
\startdata
0.12 & 0.9 & 6.0 & 14.0 & \teff~$< 6500$~K \\
0.4 & 0.8 & 20.0 & 12.75 & \teff~$> 7000$~K
\enddata
\end{deluxetable}

During the preparation of this manuscript, we were made aware of a more detailed \Gaia DR3 radial velocity error model based on the derived stellar population.\footnote{\url{https://www.cosmos.esa.int/web/gaia/science-performance}} Since our synthetic survey does not include the stellar evolutionary stage, we opt for the simple recipe that assigns errors based on the effective temperature of the stars. The coefficients provided are not identical to those adopted here, but a comparison between the two indicates that our adopted error modeling is roughly consistent with those from the more detailed model. For the most part, our error modeling falls on the conservative side of the latest model.

\cite{gaia_rv} noted that during scientific validation of the published DR3 radial velocities, the above uncertainties were underestimated and thus require an additional multiplicative correction factor $f$.
This multiplicative factor ($f$) is a function of $G_{\rm{RVS}}$,
\begin{equation}
\begin{aligned}
    f = a + b \, G_{\mathrm{RVS}} + c \, G_{\mathrm{RVS}}^2,
\end{aligned}
\end{equation}
with coefficients given in Table~\ref{tab:rv_err_correct_coeff}.
The velocity uncertainties should therefore be $f \times \sigma_{V_\mathrm{R}}$.
We note that the relation is only valid for $G_{\mathrm{RVS}} > 8$. For $G_{\mathrm{RVS}} < 8$, we still apply the correction function but assume $G_{\mathrm{RVS}} = 8$.
The correction factor is not applied directly to the uncertainties in the final \Gaia DR3 dataset.
Following that practice, we calculate this correction factor and provide it separately in the final synthetic survey.

\begin{deluxetable}{cccc}
\caption{Coefficients for $f$ as a function of $G_{\text{RVS}}$}
\label{tab:rv_err_correct_coeff}
\tablehead{
\colhead{$a$} & 
\colhead{$b$} & 
\colhead{$c$} & 
\colhead{Applicable range} 
}
\startdata
0.318 & 0.3884 & -0.02778 & $G_{\text{RVS}} < 12$~mag \\
16.554 & -2.4899 & 0.09933 & $G_{\text{RVS}} > 12$~mag
\enddata
\end{deluxetable}

\subsection{Selection Function and Data Release}
\label{sec:selectionfunction}

With error-convolved values computed, we next apply the selection function to produce the final synthetic surveys. We apply two selection functions, one for selecting stars that are detectable in all three photometric bands and another for selecting stars with reported radial velocity. 

We apply a $G$-band magnitude cut to select stars with reported photometry in each catalog.
We note that the cuts are applied on the error-convolved observed magnitudes.
We select the sample of stars with reported photometry via a cut on the observed $G$ magnitude, $3 < G_{\rm{obs}} < 21$.
This is the same selection cut applied in S20. 

To select the sample of stars with reported radial velocities, we make a cut on effective temperature, $T_{\rm eff}$ and $G_{\rm RVS}$. S20 reported radial velocity measurements for bright stars with $G_\mathrm{RVS} < 14$ and effective temperature of $3550 < T_\mathrm{eff} < 6900~\mathrm{K}$.
We extend the radial velocity selection to $3600 < T_\mathrm{eff} < 14500~\mathrm{K}$ for bright stars ($G_\mathrm{RVS} \leq 12$) and $3100 < T_\mathrm{eff} < 6750~\mathrm{K}$ for fainter stars ($12 < G_\mathrm{RVS} < 14$), in order to match the temperature range reported in \cite{gaia_rv}, reflecting the improvements from \Gaia DR2 to DR3. 

We bin the stars in each catalog by their LSR-centric distance into 10 radial slices.  
Table~\ref{tab:count} shows the total number of stars, as well as the number of stars with radial velocity measurements, in each radial slice and catalog.

\section{Results}
\label{sec:results}

\subsection{Comparison with Ananke DR2}

We compare our final synthetic survey for \Gaia DR3 using \Fire with the synthetic \Gaia DR2 survey from S20.
We updated the photometry to be consistent with \Gaia DR3, using isochrones and extinction laws corresponding to the \Gaia DR3 photometric system. 
We also updated the error modeling for photometric measurements and radial velocity measurements. 

The detailed numbers of sources in each radial bin of each galaxy are given in Table~\ref{tab:count}.
In general, there is a small increase in the total number of observed stars in the DR3 catalogs as compared to the DR2 catalogs. 
The number of stars with radial velocity measurements in each catalog has increased by $\sim 2-3$ times, as expected from the wider range of effective temperature $T_\mathrm{eff}$ in the selection cut (see Section~\ref{sec:selectionfunction}).
However, this is a more moderate increase than the factor of $5$ between the two \Gaia data releases (from $\sim 7$~millions stars in \Gaia DR2 to $\sim 33$ ~million stars in \Gaia DR3) \citep{gaia_rv}. 
This is due to the radial velocity selection cut in S20 being overly optimistic, already at $G_\mathrm{RVS} < 14$ when considering the actual performance of \Gaia~DR2 at $G_\mathrm{RVS} < 12.5$ \citep{2019A&A...622A.205K}. 
In all catalogs, the overall fraction of stars with radial velocity measurement compared to the total sample, which can be calculated from Table~\ref{tab:count}, is about $2-3\%$, which is indeed comparable to that of \Gaia DR3, which was about $\sim 2\%$ \citep{gaia_rv}.
For reference, the fractions of stars with radial velocity measurements in the \Ananke DR2 catalogs are about $1-1.5\%$.

\subsection{Synthetic Surveys} 

\begin{figure}
    \centering
    \includegraphics[width=0.48\textwidth]{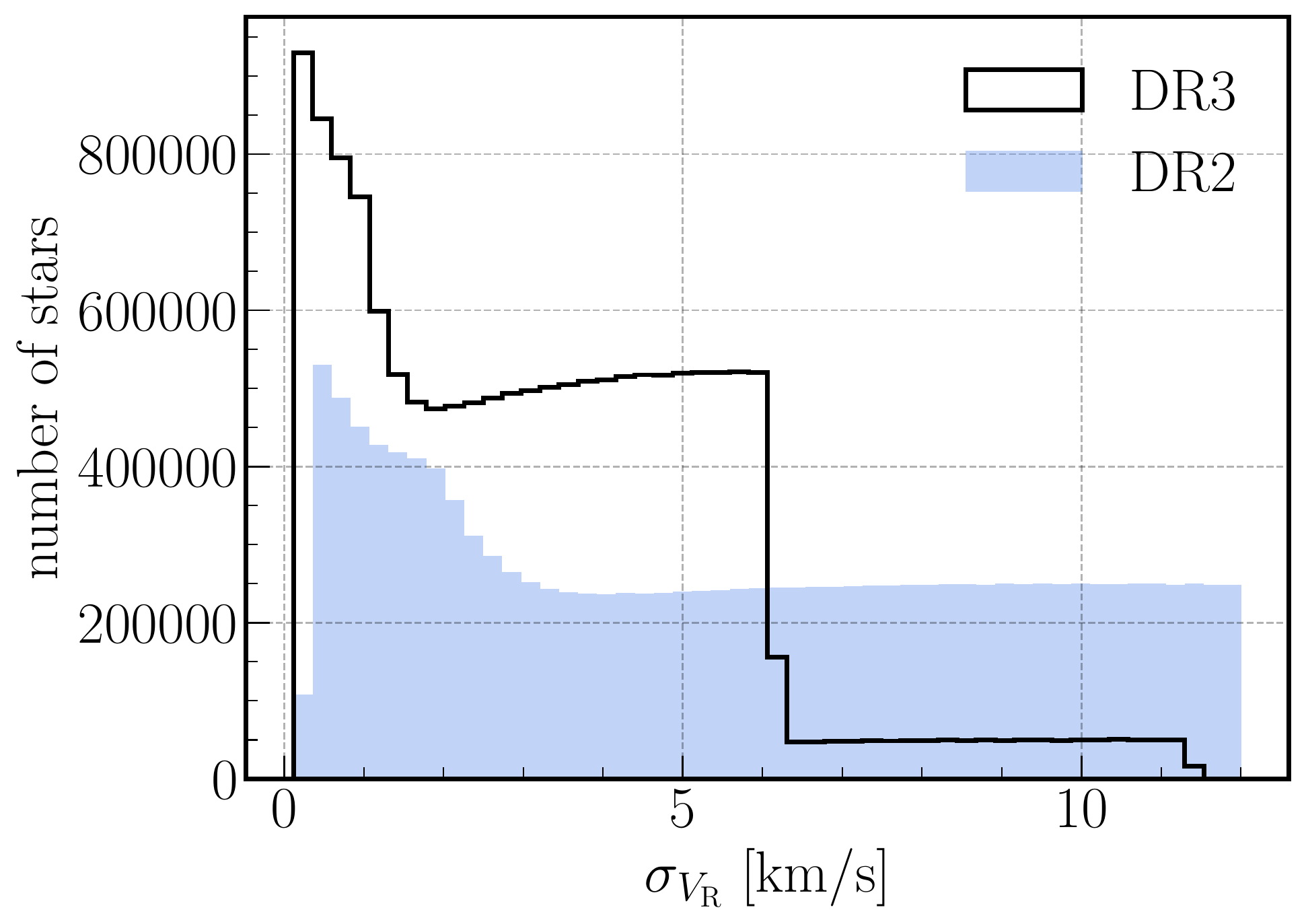}
    \caption{Distributions of the radial velocity errors for DR3 (solid black) and DR2 (blue) for all stars with radial velocities in \texttt{m12i-lsr0-rslice0}.}
    \label{fig:rv_error}
\end{figure}

\begin{figure*}
    \centering
    \includegraphics[width=0.95\textwidth]{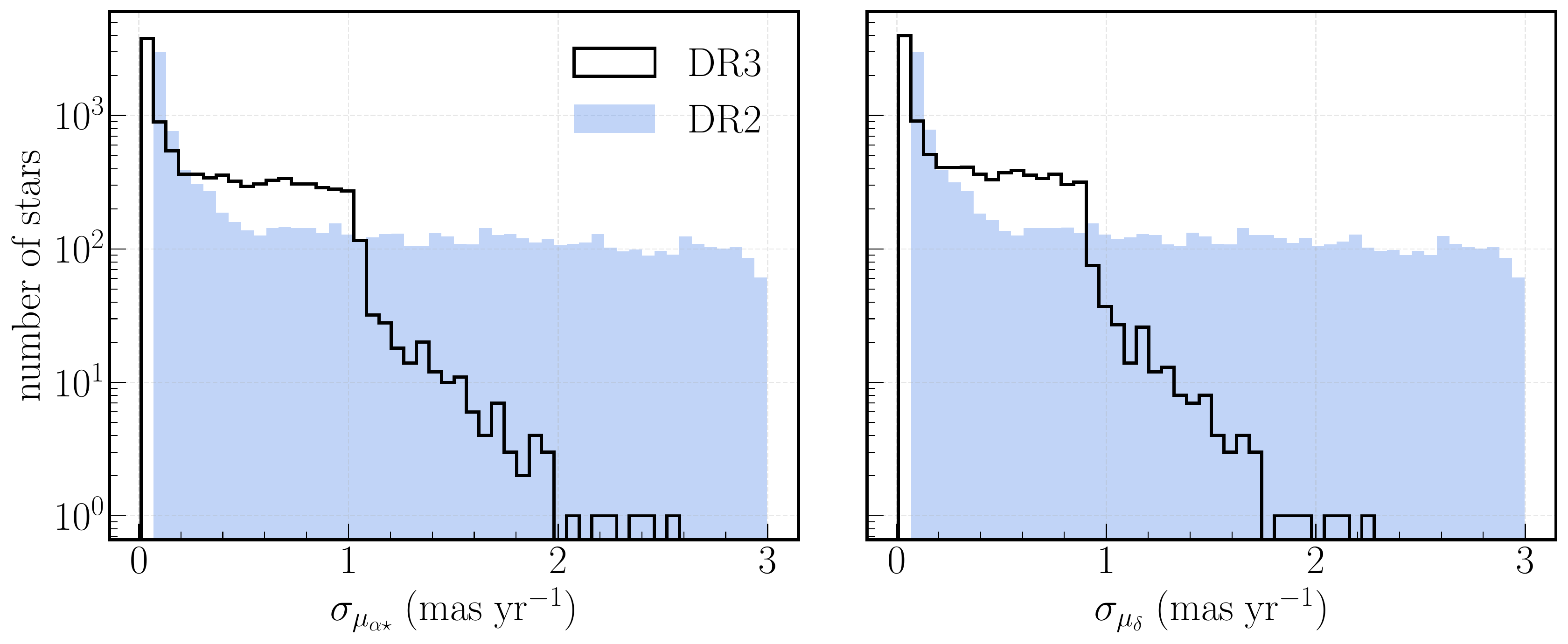}
    \caption{Distributions of the proper motion errors for RA cos(Dec) $\alpha^\star$ (left) and Dec $\delta$ (right) for DR3 (solid black) and DR2 (blue) for all stars in \texttt{m12i-lsr0-rslice0}.}
    \label{fig:pm_error}
\end{figure*}

In Figure~\ref{fig:rv_error} and Figure~\ref{fig:pm_error}, we compare the distributions of radial velocity errors and proper motion errors between DR2 and DR3 for all stars in \texttt{m12i-lsr0-rslice0}\footnote{The choice of the m12i and lsr-0 synthetic survey is just an example that we use to illustrate different properties. Similar treatment can be done with any of the other synthetic survey.}.
Figure~\ref{fig:rv_error} shows the distributions of the radial velocity errors for DR2 (blue) and DR3 (solid black) for all stars with radial velocities. 
As expected, the radial velocity errors in \Ananke DR3 are significantly lower than in DR2.
The radial velocity errors for DR3 are composed of two stellar populations: one with low \teff and one with high \teff, while the radial velocity errors in DR2 are modeled by a single exponential \citep{sanderson20}.
In \Ananke DR3, the low \teff population makes up most of the distribution below $\sigma_{V_\mathrm{R}} \lesssim 6 \; \mathrm{km/s}$, while the high \teff population is responsible for the tail at high $\sigma_{V_\mathrm{R}} \gtrsim 6 \; \mathrm{km/s}.$
The sharp cut at the lower end of the DR2 error is the systematic noise floor at $0.11 \; \mathrm{km /s}$ mentioned in S20.
Figure~\ref{fig:pm_error} shows the distributions of errors in the proper motions $\mu_{\alpha, \star}$ and $\mu_{\delta}$ for DR2 (blue) and DR3 (solid black).
Similarly, as with the radial velocity errors, the proper motion errors in \Ananke DR3, shown in Fig.~\ref{fig:pm_error} are typically much lower than in DR2. 
The DR2 proper motion errors have a cut off of $0.0861852 \; \mathrm{mas/yr}$ at the low end, as described in Equation 16 and Table 5 of S20.

\begin{figure*}
\begin{center}
\includegraphics[angle=0,width=0.85\textwidth]{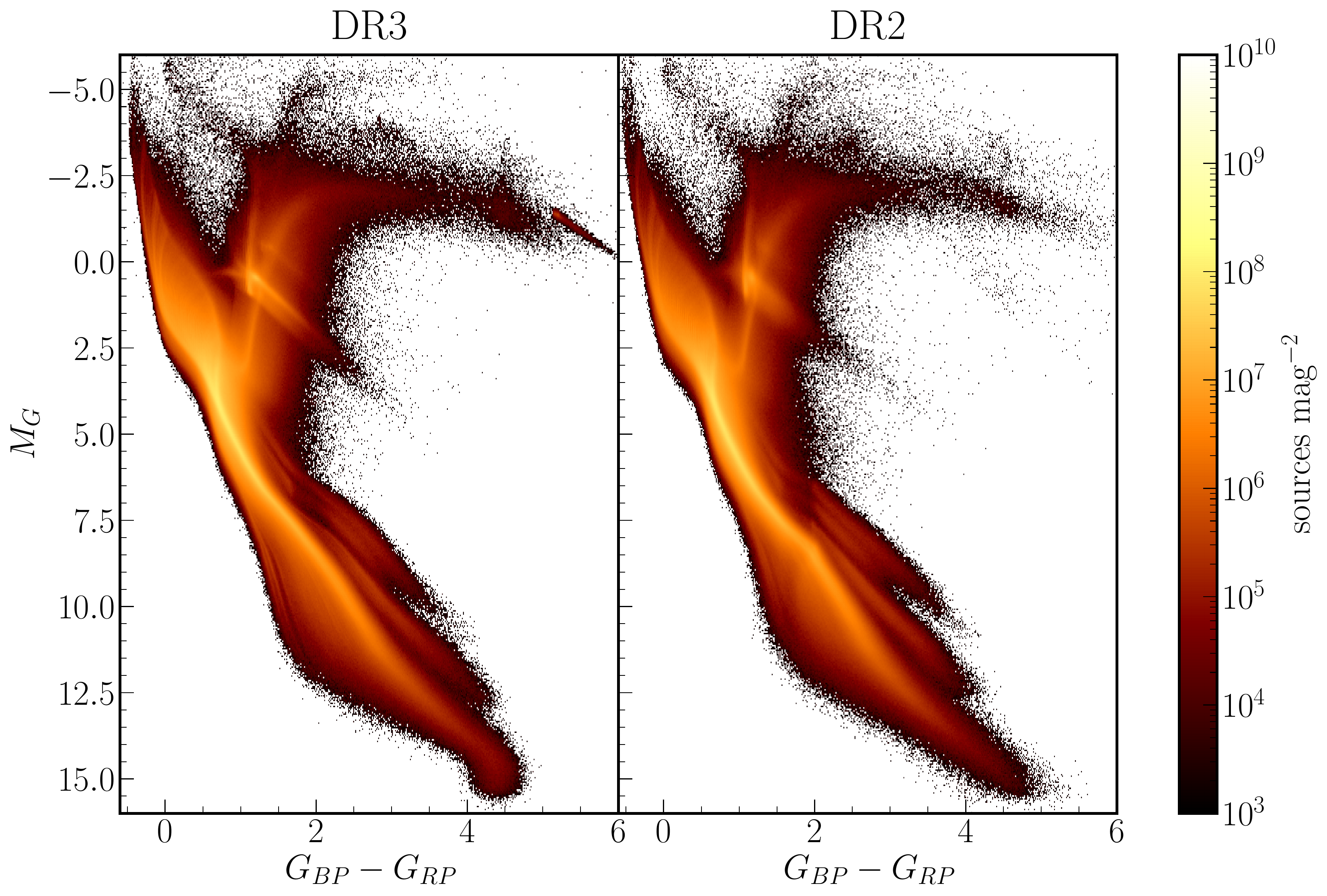}
\end{center}
\caption{The Hertzsprung–Russell diagram for \texttt{m12i-lsr0-rslice0} for stars satisfying the parallax cut $\sigma_\varpi / \varpi > 10$.}
\label{fig:hrd}
\end{figure*}

We examine the Hertzsprung–Russell 
diagram of one synthetic survey (m12i at lsr-0 rslice 0) as shown in Figure~\ref{fig:hrd}.
We plot only stars with estimated parallax error less than 10\%. 
Figure~\ref{fig:hrd} shows that our results are qualitatively similar to what was shown in \Ananke DR2 from S20.
Some echoes of the underlying grid of isochrones are still visible at the brightest magnitudes, where the model grid is sparsest, and potential artifacts from linear isochrone interpolation near the tip of the red giant branch are present (see Section~\ref{sec:synthetic_survey}). 

\begin{figure*}
\begin{center}
\includegraphics[angle=0,width=0.95\textwidth]{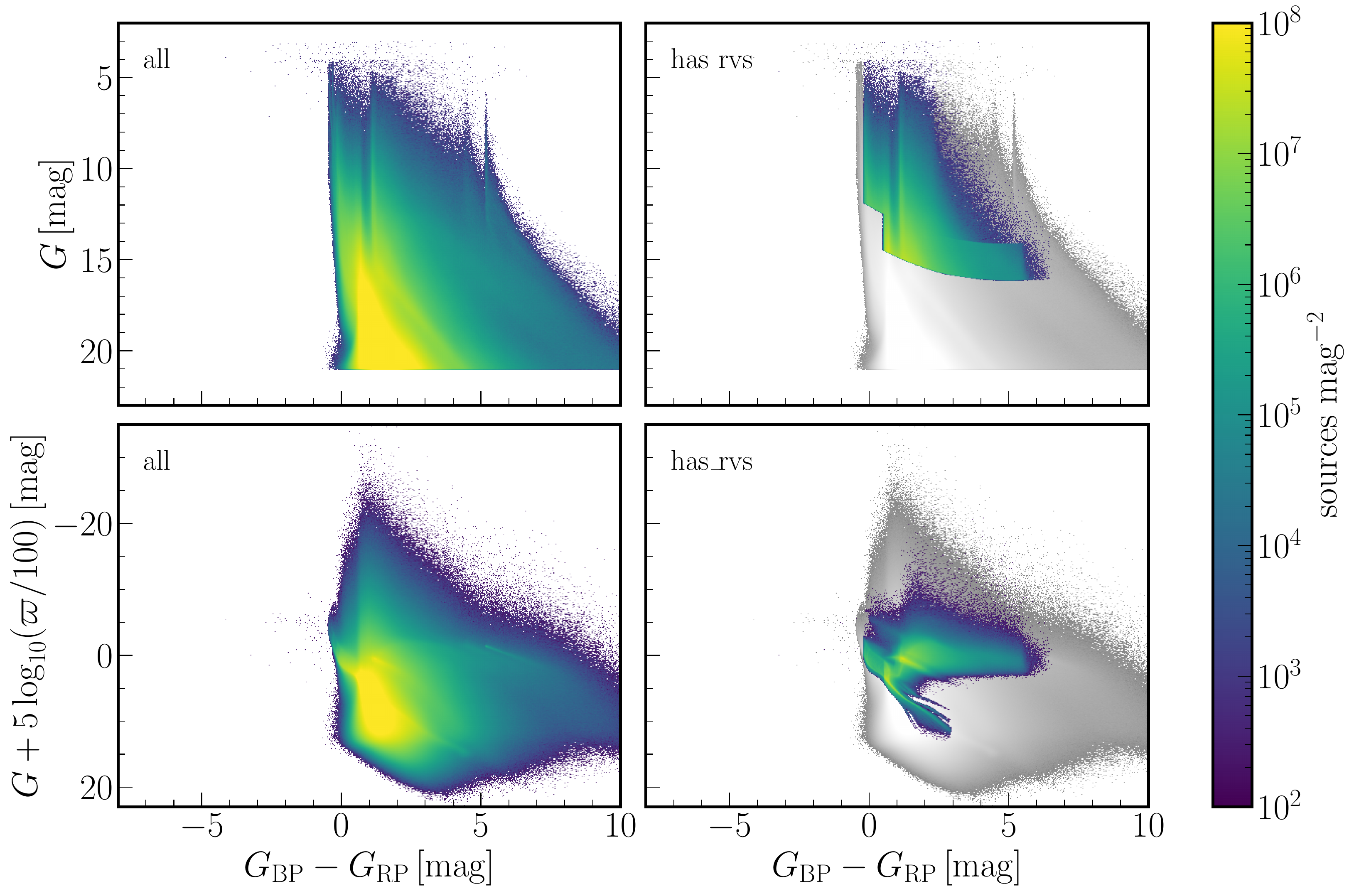}
\end{center}
\caption{The color-magnitude diagram for \texttt{m12i-lsr0-rslice0} to \texttt{m12i-lsr0-rslice9} for 
all stars (left) and those with RV measurements (right) satisfying the positive parallax cut $\varpi > 0$. The top panels shows the observed color–magnitude diagram, whereas the bottom panels shows the absolute $G$ magnitude computed from the measured parallax.}
\label{fig:cmd}
\end{figure*}

We additionally compare our results with an actual \Gaia DR3 CMD from \citet{fouesneau22}. In Figure~\ref{fig:cmd}, we partially reproduce Figure~1 in \citet{fouesneau22}. 
Our synthetic survey generates CMDs qualitatively similar to the \Gaia DR3 data. 
When we only consider the subsample with RV measurements (\texttt{has\_rvs}), the synthetic survey distributions qualitatively resemble that of the real \Gaia DR3 survey.

\subsection{List of parameters in the synthetic surveys}
\label{sec:column_list}

In Table~\ref{tab:data_model}, we present the column names of the parameters used in the synthetic surveys, as well as their definitions, data types, and units. These column names are categorized by those matching \Gaia~DR2/DR3 (as well as \Ananke~DR2), those that are relating the properties of the simulations (for example the true non-error convolved values), and the properties of the stars (for example their \Fire-2 chemical abundances). 

\section{Use Cases and Limitations}
\label{sec:limitations}

Synthetic surveys can be extremely powerful in testing modeling procedures, calculating false positive rates, and validating methods. This is largely due to the fact that cosmological simulations in general track non-equilibrium dynamics self-consistently, and are therefore powerful tools for exploring dynamical inferences. The need for synthetic surveys is becoming even more critical with the large swaths of data being collected by current and upcoming surveys like \Gaia and LSST.  There are, however, limitations to these studies, based on the nature of the construction of the synthetic surveys. In particular, as was highlighted above, the original simulations contained star particles of masses $\sim 7000 M_{\odot}$, from which we spawn individual stars. The resulting positions and velocities of the synthetic stars depend on the kernel of choice, limiting the usage of such synthetic surveys. In particular, the internal dynamics of small-scale structures like satellite galaxies and stellar streams are sensitive to the choice of kernel. Studies of small-scale MW structures therefore required careful kernel selection based on the science question at hand; \cite{Shipp:2023}, for example, changed the kernel to be able to perform detectability studies of stellar streams with LSST. 

More generally, studies of large stellar structures (larger than a few star particles in the original simulation), velocity anisotropies of the Galaxy, and the MW potential, should be robust to the choices of construction of the synthetic surveys, while studies of smaller structures should be treated with care, and potentially a more adequate choice of kernels.  

Additionally, the original simulations from which we built these synthetic surveys are not meant to reproduce the MW itself. Indeed, these are cosmological simulations with varying initial conditions, and therefore varying histories; some contain a late merger like m12f, while m12i has a quieter merger history, for example \citep[see e.g.][for details of the merger histories of these two galaxies]{2019ApJ...883...27N}. Therefore, it is critical to treat these galaxies as examples of galaxies with the same mass as our own, but with their own individual properties. Corollary to this, the dust model adapted for these synthetic surveys is self-consistent with that of the simulations themselves, and therefore is different from that of the MW. 

\section{Conclusions}

In this paper, we presented a new set of synthetic surveys that match \Gaia DR3, based on the \Latte suite of the \Fire-2 simulations. This is an update to the synthetic surveys released by \cite{sanderson20} that matched the previous data release \Gaia DR2. These synthetic surveys include three different solar positions for three galaxies. The major changes compared to S20 are an updated set of isochrones matching the latest release, a different treatment of the radial velocity errors that increased the precision of the radial velocity measurements, an update to the proper motion treatment, which decreased the measurement proper motion errors in the synthetic surveys, and an increase in the total number of stars with radial velocity measurements through the update of the selection cuts.

These synthetic surveys are made available to the community on \url{http://ananke.hub.yt/}, where they can be used to test any model/analysis pipeline on simulations prior to application to \Gaia DR3. In particular, these synthetic surveys are the best tool for studies involving the dynamic properties of the MW, especially given that the ``true" star particles from the original simulations are also provided.

More generally, the adoption of synthetic surveys is not only applicable in the study of stars, but also the dynamics of dark matter and properties of gas particles. \cite{2023ApJS..265...44W} has made the simulations used in this work publicly available\footnote{\url{http://flathub.flatironinstitute.org/fire}}, including the formation coordinates of all star particles, as well as catalogs of all satellite galaxies/halos. Therefore, the community can use such information to answer more general questions as to what the field can learn through \Gaia about the galaxy as a whole, from its stellar components as measured through the \Gaia lens, to the inner workings of the dark matter in the Galaxy that governs the dynamics of the stars. 

\section*{Acknowledgements}
We thank Kacper Kowalik and Matthew Turk for their tremendous help on data storage. We also thank Sarah Loebman for her work on the original \Ananke paper. 

AW received support from: NSF via CAREER award AST-2045928 and grant AST-2107772; NASA ATP grant 80NSSC20K0513; HST grants AR-15809, GO-15902, GO-16273 from STScI.

NP and RES acknowledge support from NASA grant 19-ATP19-0068. NP was supported in part by a Zacheus Daniel fellowship from the University of Pennsylvania. RES additionally acknowledges support from  NSF grant AST-2007232, from the Research Corporation through the Scialog Fellows program on Time Domain Astronomy, and from HST-AR-15809 from the Space Telescope Science Institute (STScI), which is operated by AURA, Inc., under NASA contract NAS5-26555. 

This work made use of Stampede-2, a large computing cluster funded by the National Science Foundation (NSF) through award ACI-1540931. The analysis was conducted using the allocation number PHY210118.

This research or product makes use of public auxiliary data provided by ESA/Gaia/DPAC/CU5 and prepared by Carine Babusiaux. This work has made use of data from the European Space Agency (ESA) mission
{\it Gaia} (\url{https://www.cosmos.esa.int/gaia}), processed by the {\it Gaia}
Data Processing and Analysis Consortium (DPAC,
\url{https://www.cosmos.esa.int/web/gaia/dpac/consortium}). Funding for the DPAC
has been provided by national institutions, in particular the institutions
participating in the {\it Gaia} Multilateral Agreement.

\begin{longrotatetable}
\begin{deluxetable*}{llcc}
\label{tab:data_model}
\tabletypesize{\footnotesize}
\tablecolumns{4}
\tablecaption{Data Model for Synthetic Surveys}
\tablehead{
\colhead{Quantity} &
\colhead{Explanation} &
\colhead{Data type} &
\colhead{Unit} 
}
\startdata
\cutinhead{\textit{Fields with names identical to those in DR2}}
\cutinhead{Astrometry}
\texttt{ra} & Right ascension & double & Angle (deg) \\
\texttt{ra\_error} & Standard error of R.A. & double & Angle (deg) \\
\texttt{dec} & decl. & double & Angle (deg) \\
\texttt{dec\_error} & Standard error of decl. & double & Angle (deg) \\
\texttt{parallax} & Parallax & double & Angle (mas) \\
\texttt{parallax\_error} & Standard error of parallax & double & Angle (mas) \\
\texttt{parallax\_over\_error} & Parallax divided by its error & float & \nodata \\
\texttt{pmra} & Proper motion in R.A. direction & double & Angular Velocity (\masyr) \\
\texttt{pmra\_error} & Standard error of proper motion in R.A. direction & double & Angular Velocity (\masyr) \\
\texttt{pmdec} & Proper motion in decl. direction & double & Angular Velocity (\masyr) \\
\texttt{pmdec\_error} & Standard error of proper motion in decl. direction & double & Angular Velocity (\masyr) \\
\texttt{l} & Galactic longitude (converted from R.A., decl.) & double & Angle (deg) \\
\texttt{b} & Galactic latitude (converted from R.A., decl.) & double & Angle (deg) \\
\cutinhead{Photometry}
\texttt{phot\_g\_mean\_mag} & Extincted apparent $G$-band mean magnitude & float & Magnitude (mag) \\
\texttt{phot\_bp\_mean\_mag} & Extincted apparent $G_{BP}$-band mean magnitude & float & Magnitude (mag) \\
\texttt{phot\_rp\_mean\_mag} & Extincted apparent $G_{RP}$-band mean magnitude & float & Magnitude (mag) \\
\texttt{bp\_rp} & Reddened $G_{BP}-G_{RP}$ color & float & Magnitude (mag) \\
\texttt{bp\_g} & Reddened $G_{BP}-G$ color & float & Magnitude (mag) \\
\texttt{g\_rp} & Reddened $G-G_{RP}$ color & float & Magnitude (mag) \\
\cutinhead{Spectroscopy}
\texttt{radial\_velocity} & Radial velocity & double & Velocity (\kmsec) \\
\texttt{radial\_velocity\_error} & Standard error of radial velocity & double & Velocity (\kmsec) \\
\cutinhead{\textit{Other fields not in the Gaia DR2 data model}}
\cutinhead{Indices}
\texttt{starid} & array index of the star (per mock catalog) & long & \nodata \\
\texttt{parentid} & array index of the generating star particle in the snapshot file & long & \nodata \\
\texttt{partid} & \shortstack[l]{0 if phase-space coordinates are identical to the generating star particle,\\ 1 otherwise} & short & \nodata \\
\cutinhead{Phase Space}
\texttt{ra\_true} & true R.A. & double & Angle (deg) \\
\texttt{dec\_true} & true decl. & double & Angle (deg) \\
\texttt{dmod\_true} & true distance modulus & double & Magnitude (mag) \\
\texttt{ra\_cosdec\_error} & standard error in R.A.$\cos{(\rm decl.)}$ & double & Magnitude (deg) \\
\texttt{parallax\_true} & true parallax & double & Angle (mas) \\
\texttt{pmra\_true} & true pm in R.A. direction & double & Angular Velocity (\masyr) \\
\texttt{pmdec\_true} & true pm in decl. direction & double & Angular Velocity (\masyr) \\
\texttt{radial\_velocity\_true} & true RV & double & Velocity (\kmsec) \\
\texttt{l\_true} & true Galactic long & double & Angle (deg) \\
\texttt{b\_true} & true Galactic lat & double & Angle (deg) \\
\texttt{pml} & pm in Galactic long direction & double & Angular Velocity (\masyr) \\
\texttt{pmb} & pm in Galactic lat direction & double & Angular Velocity (\masyr) \\
\texttt{pml\_true} & true pm in Galactic long direction & double & Angular Velocity (\masyr) \\
\texttt{pmb\_true} & true pm in Galactic lat direction & double & Angular Velocity (\masyr) \\
\texttt{px\_true, py\_true, pz\_true} & true position relative to LSR & double & Distance (kpc) \\
\texttt{vx\_true, vy\_true, vz\_true} & true velocity relative to LSR & double & Velocity (\kmsec) \\
\cutinhead{Photometry}
\texttt{phot\_g\_mean\_mag\_true} & \shortstack[l]{true (i.e., after extinction, but before error convolution) apparent $G$-band\\ mean magnitude} & float & Magnitude (mag) \\
\texttt{phot\_bp\_mean\_mag\_true} & true apparent $G_{BP}$-band mean magnitude & float & Magnitude (mag) \\
\texttt{phot\_rp\_mean\_mag\_true} & true apparent $G_{RP}$-band mean magnitude & float & Magnitude (mag) \\
\texttt{phot\_g\_mean\_mag\_int} & \shortstack[l]{intrinsic (i.e., before extinction or error convolution) apparent $G$-band\\ mean magnitude} & float & Magnitude (mag) \\
\texttt{phot\_bp\_mean\_mag\_int} & intrinsic apparent $G_{BP}$-band mean magnitude & float & Magnitude (mag) \\
\texttt{phot\_rp\_mean\_mag\_int} & intrinsic apparent $G_{RP}$-band mean magnitude & float & Magnitude (mag) \\
\texttt{phot\_g\_mean\_mag\_abs} & absolute $G$-band mean magnitude & float & Magnitude (mag) \\
\texttt{phot\_bp\_mean\_mag\_abs} & absolute $G_{BP}$-band mean magnitude & float & Magnitude (mag) \\
\texttt{phot\_rp\_mean\_mag\_abs} & absolute $G_{RP}$-band mean magnitude & float & Magnitude (mag) \\
\texttt{phot\_g\_mean\_mag\_error} & Standard error of $G$-band mean magnitude & float & Magnitude (mag) \\
\texttt{phot\_bp\_mean\_mag\_error} & Standard error of $G_{BP}$-band mean magnitude & float & Magnitude (mag) \\
\texttt{phot\_rp\_mean\_mag\_error} & Standard error of $G_{RP}$-band mean magnitude & float & Magnitude (mag) \\
\texttt{bp\_rp\_true} & true $G_{BP}-G_{RP}$ color & float & Magnitude (mag) \\
\texttt{bp\_g\_true} & true $G_{BP}-G$ color & float & Magnitude (mag) \\
\texttt{g\_rp\_true} & true $G-G_{RP}$ color & float & Magnitude (mag) \\
\texttt{vmini\_true} & true $V-I$ color used for error modeling & float & Magnitude (mag) \\
\cutinhead{Extinction}
\texttt{lognh} & $\log_{10}$ equivalent H column density along line of sight to star & float & surface number density(cm$^{\rm -2}$) \\
\texttt{ebv} & $E(B-V)$ reddening, calculated from $N_{\rm H}^{eff}$ & float & Magnitude (mag) \\
\texttt{A0} & $A_0$, extinction at 550\,nm, assuming $R_V = 3.1$ & float & Magnitude (mag) \\
\texttt{a\_g\_val} & true line-of-sight extinction in the $G$ band, $A_G$ & float & Magnitude (mag) \\
\texttt{e\_bp\_min\_rp\_val} & true line-of-sight reddening $G_{BP}-G_{RP}$ & float & Magnitude (mag) \\
\cutinhead{Spectroscopy}
\texttt{radial\_velocity\_error\_corr\_factor} & correction factor for \texttt{radial\_velocity\_error} & double & Velocity (\kmsec) \\
\cutinhead{Stellar Parameter}
\texttt{mact} & current stellar mass & float & Mass (Solar Mass) \\
\texttt{mtip} & mass of a star at tip of giant branch for given age, metallicity & float & Mass (Solar Mass) \\
\texttt{mini} & stellar mass on zero-age main sequence & float & Mass (Solar Mass) \\
\texttt{age} & $\log_{10}$ of stellar age; identical for all stars generated from the same particle & float & Time (log yr) \\
\texttt{teff} & stellar effective temperature & float & Temperature (K) \\
\texttt{logg} & surface gravity & float & Surface Gravity (log cgs) \\
\texttt{lum} & $\log_{10}$ of stellar luminosity & float &  Luminosity (log Solar Luminosity) \\
\cutinhead{Abundances}
\texttt{feh} & [Fe/H] & float & Magnitude (mag) \\
\texttt{alpha} & [Mg/Fe] & float & Magnitude (mag) \\
\texttt{carbon} & [C/H] & float & Magnitude (mag) \\
\texttt{helium} & [He/H] & float & Magnitude (mag) \\
\texttt{nitrogen} & [N/H] & float & Magnitude (mag) \\
\texttt{sulphur} & [S/H] & float & Magnitude (mag) \\
\texttt{oxygen} & [O/H] & float & Magnitude (mag) \\
\texttt{silicon} & [Si/H] & float & Magnitude (mag) \\
\texttt{calcium} & [Ca/H] & float & Magnitude (mag) \\
\texttt{magnesium} & [Mg/H] & float & Magnitude (mag) \\
\texttt{neon} & [Ne/H] & float & Magnitude (mag) \\
\cutinhead{Quality Control}
\texttt{flag\_wd} & flag for potential white dwarfs; see Section~\ref{sec:isochrones} & int & \nodata 
\enddata
\end{deluxetable*}
\end{longrotatetable}

\begin{table*}
\caption{Number of stars in the \Ananke DR3 surveys of the \Latte
MW-mass suite of FIRE simulations.}
\label{tab:count}
\centering
\begin{tabularx}{\linewidth}{III|IIIIII}
\hline \hline
\multicolumn{3}{c|}{\multirow{2}{*}{File Information}} & \multicolumn{6}{c}{\multirow{2}{*}{Number of stars}} \\
\multicolumn{3}{c|}{} & \multicolumn{6}{c}{} \\ \hline
\multicolumn{1}{l}{} & $d_\mathrm{min}$ & $d_\mathrm{max}$ & \multicolumn{3}{c|}{m12i} & \multicolumn{3}{c}{m12i with radial velocity} \\
index & [kpc] & [kpc] & lsr-0 & lsr-1 & \multicolumn{1}{c|}{lsr-2} & lsr-0 & lsr-1 & lsr-2 \\ \hline
0 & 0 & 3 & $316,095,707$ & $241,317,358$ & \multicolumn{1}{c|}{$262,072,171$} & $15,646,316$ & $18,034,724$ & $19,356,555$ \\
1 & 3 & 4.25 & $290,904,524$ & $221,518,272$ & \multicolumn{1}{c|}{$213,845,515$} & $5,468,516$ & $5,526,892$ & $5,503,211$ \\
2 & 4.25 & 5.5 & $401,479,587$ & $296,537,911$ & \multicolumn{1}{c|}{$245,927,828$} & $7,015,408$ & $7,420,804$ & $6,441,700$ \\
3 & 5.5 & 6.5 & $400,845,878$ & $292,297,437$ & \multicolumn{1}{c|}{$213,285,291$} & $6,624,552$ & $7,427,973$ & $5,943,614$ \\
4 & 6.5 & 7.25 & $365,130,175$ & $261,197,349$ & \multicolumn{1}{c|}{$168,853,179$} & $6,257,625$ & $6,771,239$ & $5,393,689$ \\
5 & 7.25 & 8 & $418,818,886$ & $291,579,170$ & \multicolumn{1}{c|}{$172,874,861$} & $8,324,411$ & $8,044,724$ & $6,679,657$ \\
6 & 8 & 9 & $507,799,164$ & $338,792,594$ & \multicolumn{1}{c|}{$198,177,685$} & $11,999,243$ & $10,649,686$ & $9,100,738$ \\
7 & 9 & 10 & $320,749,442$ & $212,312,749$ & \multicolumn{1}{c|}{$127,489,961$} & $8,105,168$ & $7,231,158$ & $6,285,887$ \\
8 & 10 & 15 & $510,906,338$ & $340,848,426$ & \multicolumn{1}{c|}{$227,453,144$} & $15,677,739$ & $13,843,185$ & $12,735,550$ \\
9 & 15 & 300 & $149,436,821$ & $75,879,230$ & \multicolumn{1}{c|}{$68,095,443$} & $11,557,428$ & $9,579,237$ & $8,934,000$ \\ \hline
\multicolumn{1}{l}{} & \multicolumn{1}{l}{} & Total & $3,682,166,522$ & $2,572,280,496$ & \multicolumn{1}{c|}{$1,898,075,078$} & $96,676,406$ & $94,529,622$ & $86,374,601$ \\
\multicolumn{1}{l}{} & \multicolumn{1}{l}{} & DR2 Total & $3,215,565,725$ & $3,754,501,977$ & \multicolumn{1}{c|}{$2,932,162,112$} & $38,183,839$ & $44,583,007$ & $39,191,496$ \\ \hline \hline
\multicolumn{1}{l}{} & $d_\mathrm{min}$ & $d_\mathrm{max}$ & \multicolumn{3}{c|}{m12f} & \multicolumn{3}{c}{m12f with radial velocity} \\
index & [kpc] & [kpc] & lsr-0 & lsr-1 & \multicolumn{1}{c|}{lsr-2} & lsr-0 & lsr-1 & lsr-2 \\ \hline
0 & 0 & 3 & $295,175,898$ & $405,797,970$ & \multicolumn{1}{c|}{$376,770,540$} & $20,157,602$ & $30,374,955$ & $26,192,624$ \\
1 & 3 & 4.25 & $309,565,802$ & $305,803,461$ & \multicolumn{1}{c|}{$321,184,724$} & $7,536,353$ & $7,838,772$ & $8,082,358$ \\
2 & 4.25 & 5.5 & $442,874,027$ & $368,227,799$ & \multicolumn{1}{c|}{$360,338,273$} & $11,024,894$ & $9,865,787$ & $9,407,981$ \\
3 & 5.5 & 6.5 & $436,280,986$ & $329,184,459$ & \multicolumn{1}{c|}{$313,988,725$} & $11,584,645$ & $9,114,645$ & $8,932,140$ \\
4 & 6.5 & 7.25 & $394,454,859$ & $272,363,497$ & \multicolumn{1}{c|}{$257,757,420$} & $11,274,029$ & $8,714,783$ & $8,673,940$ \\
5 & 7.25 & 8 & $455,237,384$ & $305,888,428$ & \multicolumn{1}{c|}{$287,708,155$} & $13,781,431$ & $12,009,745$ & $11,436,403$ \\
6 & 8 & 9 & $544,181,654$ & $362,325,501$ & \multicolumn{1}{c|}{$346,300,118$} & $17,645,803$ & $17,165,788$ & $15,961,569$ \\
7 & 9 & 10 & $351,832,549$ & $210,996,318$ & \multicolumn{1}{c|}{$218,765,182$} & $11,106,948$ & $10,507,034$ & $10,193,449$ \\
8 & 10 & 15 & $708,763,955$ & $374,315,091$ & \multicolumn{1}{c|}{$448,965,495$} & $19,482,838$ & $17,556,943$ & $17,498,773$ \\
9 & 15 & 300 & $327,448,929$ & $184,948,854$ & \multicolumn{1}{c|}{$275,426,823$} & $14,942,176$ & $12,791,740$ & $14,172,354$ \\ \hline
\multicolumn{1}{l}{} & \multicolumn{1}{l}{} & Total & $4,265,816,043$ & $3,119,851,378$ & \multicolumn{1}{c|}{$3,207,205,455$} & $138,536,719$ & $135,940,192$ & $130,551,591$ \\
\multicolumn{1}{l}{} & \multicolumn{1}{l}{} & DR2 Total & $5,851,407,276$ & $4,706,540,756$ & \multicolumn{1}{c|}{$4,678,842,172$} & $62,673,864$ & $61,393,185$ & $57,808,862$ \\ \hline \hline
\multicolumn{1}{l}{} & $d_\mathrm{min}$ & $d_\mathrm{max}$ & \multicolumn{3}{c|}{m12m} & \multicolumn{3}{c}{m12m with radial velocity} \\
index & [kpc] & [kpc] & lsr-0 & lsr-1 & \multicolumn{1}{c|}{lsr-2} & lsr-0 & lsr-1 & lsr-2 \\ \hline
0 & 0 & 3 & $984,809,951$ & $1,073,978,992$ & \multicolumn{1}{c|}{$910,734,608$} & $47,393,328$ & $54,240,600$ &  $43,119,764$\\
1 & 3 & 4.25 & $728,265,777$ & $798,150,011$ & \multicolumn{1}{c|}{$686,462,276$} & $12,171,491$ & $13,592,305$ & $11,429,030$ \\
2 & 4.25 & 5.5 & $814,806,044$ & $863,540,944$ & \multicolumn{1}{c|}{$796,780,191$} & $12,767,460$ & $13,727,680$ & $12,465,994$ \\
3 & 5.5 & 6.5 & $685,954,361$ & $723,050,215$ & \multicolumn{1}{c|}{$689,642,062$} & $10,229,572$ & $10,706,710$ & $9,751,780$ \\
4 & 6.5 & 7.25 & $528,436,556$ & $558,951,415$ & \multicolumn{1}{c|}{$531,816,221$} & $7,774,165$ & $7,855,599$ & $7,300,150$ \\
5 & 7.25 & 8 & $523,399,484$ & $551,230,847$ & \multicolumn{1}{c|}{$532,527,598$} & $7,611,444$ & $7,589,807$ & $7,471,248$ \\
6 & 8 & 9 & $2,003,093,353$ & $639,727,826$ & \multicolumn{1}{c|}{$617,194,075$} & $17,707,039$ & $9,881,911$ & $10,009,474$ \\
7 & 9 & 10 & $422,716,282$ & $458,827,031$ & \multicolumn{1}{c|}{$432,726,049$} & $8,073,088$ & $8,469,156$ & $8,438,569$ \\
8 & 10 & 15 & $835,507,954$ & $1,267,343,926$ & \multicolumn{1}{c|}{$1,192,679,933$} & $23,167,408$ & $26,128,219$ & $26,126,666$ \\
9 & 15 & 300 & $261,056,409$ & $268,075,320$ & \multicolumn{1}{c|}{$244,124,886$} & $20,074,951$ & $20,492,819$ & $19,063,333$ \\ \hline
\multicolumn{1}{l}{} & \multicolumn{1}{l}{} & Total & $7,788,046,171$ & $7,202,876,527$ & \multicolumn{1}{c|}{$6,634,687,899$} & $166,969,946$ & $172,684,806$ & $155,176,008$ \\
\multicolumn{1}{l}{} & \multicolumn{1}{l}{} & DR2 Total & $5,701,759,381$ & $6,415,674,623$ & \multicolumn{1}{c|}{$5,516,835,110$} & $84,931,532$ & $108,808,464$ & $78,520,886$ \\ \hline \hline
\end{tabularx}
\end{table*}

\bibliography{sample631}
\bibliographystyle{aasjournal}

\end{document}